\documentclass[11pt]{amsart}

\usepackage{amsmath,amssymb,amsthm,amsfonts,graphicx,fullpage}%
\usepackage{booktabs}%
\usepackage[tableposition=below]{caption}%
\usepackage{subfigure}
\usepackage{mathpple}
\usepackage{hyperref,enumerate}

\pdfoutput=1

\usepackage{colortbl}
\usepackage[all,cmtip]{xy}

\newcommand{\bra}[1]{\langle #1 |}
\newcommand{\ket}[1]{| #1 \rangle}
\def\a{a_{\textup{FLRW}}} 
\def\ahg{\mathfrak{a}} 
\def\adot{\dot{a}_{\textup{FLRW}}}
\def\addot{\ddot{a}_{\textup{FLRW}}}
\def\bhg{\mathfrak{b}} 
\def\bp{b} 
\def\chg{\mathfrak{c}} 
\def\O{\mathcal{O}}
\def\p{\mathbf p}
\def\R{\mathbb{R}}
\def\x{\mathbf x}

\usepackage[sort&compress,numbers]{natbib}

\title{Particle Creation in a Cosmological Background in Analogy to the Schwinger Effect}
\date{May 12, 2026\\
Walter van Suijlekom and Michael Wondrak as the first authorship.}

\author{Walter D. van Suijlekom}
\email{waltervs@math.ru.nl}
\address{Institute for Mathematics, Astrophysics and Particle Physics (IMAPP), Radboud University, P.O.\ Box 9010,
6500 GL Nijmegen, The Netherlands}

\author{Michael F. Wondrak}
\email{m.wondrak@astro.ru.nl}
\address{Institute for Mathematics, Astrophysics and Particle Physics (IMAPP), Radboud University, P.O.\ Box 9010,
6500 GL Nijmegen, The Netherlands}
\address{Anton Pannekoek Institute for Astronomy, University of Amsterdam, P.O.\ Box 94249, 1090 GE Amsterdam, The Netherlands}
\address{Gravitation and Astroparticle Physics Amsterdam (GRAPPA), University of Amsterdam, Science Park 904, 1098 XH Amsterdam, The Netherlands}

\author{Heino Falcke}
\email{h.falcke@astro.ru.nl}
\address{Institute for Mathematics, Astrophysics and Particle Physics (IMAPP), Radboud University, P.O.\ Box 9010,
6500 GL Nijmegen, The Netherlands}

\begin{document}

\begin{abstract}
We consider a gravitational analogue of the Schwinger effect in a cosmological context. While the Schwinger effect is usually attributed to a static electric background, its derivation is actually based on a switching on/off of the electric field in the infinite past/future. Motivated by this, and our previous work on particle production in a gravitational background, we consider a long pulse of the gravitational field in an FLRW-spacetime, thus simulating a static background. We rigorously derive particle production by a novel application of the Heun equation. In fact, the recently obtained connecting formulas between its local solutions can be used to determine the Bogolyubov coefficients, and subsequently the particle production probabilities, in the limit of an infinitely long pulse. 

The particle production in the FLRW-model is found to have a lower threshold on the late time frequencies, which can be related to the duration of the time interval of the switching on/off the background field. For large frequencies and in the spatially flat case, we find a Planckian frequency spectrum whose temperature is inversely proportional to the duration of scale change. 

We compare our findings to Schwinger's result on particle production for a long pulse of an electromagnetic background field, for which we also include a detailed derivation.  
           
\end{abstract}

\maketitle
\tableofcontents

\section{Introduction}
There is a striking conceptual similarity between the Schwinger effect and Hawking radiation: where the former is a mechanism of particle production in a constant electromagnetic background field, the latter describes particle production in a gravitational background. This has been studied at length in the literature; see \cite{GMM94,BMPS95} for reviews. 
In our previous work \cite{WSF23,FWS25}, we also built on this analogy, basing ourselves on heat kernel methods.
The derivation of the effective Lagrangian therein was based on a Wick rotation to the Euclidean section of the spacetime manifold, thus mathematically justifying the use of heat-kernel methods. Early works along these lines are \cite{ZS77} and \cite{Faw83}. A crucial point that should be stressed is that the Euclidean section is periodic in time, which corresponds to the consideration of a thermal state in Lorentzian spacetime \cite{Wal84}, {\em i.e.} the so-called Hartle--Hawking state. A variation on the exploited methods yields both supporting and contrasting results; see, for instance, \cite{FNP23,Che23,ADS2024,BFM24,WSF24} for recent and ongoing discussions.

This paper is an attempt to shed further light on the origin of particle production in both an electromagnetic and a gravitational background field. Schwinger derived particle production for a switch on and off of an electric field pulse, considering the limit where the length of the pulse interval goes to infinity. Motivated by this, we consider an FLRW-model where the gravitational background is switched on and off, and then consider the analogous limit of a long pulse.

We point out the similarities and the contrasts between the electromagnetic and gravitational case. We perform our analysis in the rigorous context of canonical quantization of free scalar fields in the pertinent backgrounds. In the limit of a long pulse of these background fields, we consider the Bogolyubov transformations relating early time modes to late time modes. In particular, we identify the final modes obtained by such transformations when applied to an initial vacuum state.  Such modes are then understood as the potential source of particle production. For a related study based on the phase integral approximation method and numerics using a different time profile, see \cite{AMS18}.

Even though the derivation of the Bogolyubov transformations and the analysis of the probability of particle production may be technically somewhat involved, at the conceptual level the derivation seems less complicated. In fact, in all cases, the recipe is as follows.
\begin{enumerate}
\item Derive the field equation for a quantum scalar field in a background (gravitational/electric) field which is turned on during a certain (long) time interval;
\item Solve the field equation for modes that have positive frequency at early times;
\item Derive from the connecting formulas for the local solutions of the field equations the Bogolyubov transformations relating early time modes to late time modes; the latter may be both positive and negative frequency modes;
\item In the Heisenberg picture, consider the vacuum for the early time observables;
\item Using the Bogolyubov transformations, analyze the particle content for this vacuum state using late time observables;
\item Consider a field pulse in the limit of an infinitely long time interval, which thus becomes a constant field background;
\item Interpret the particle creation as being due to the constant field background and/or its switching on/off.
\end{enumerate}
We apply this in Section \ref{sect:schw} to a long electric field pulse to (re)derive the Schwinger effect \cite{Sch51,GMM94}. Section \ref{sect:flrw} then considers a novel gravitational model, which is based on a scale function that describes a gravitational pulse in an FLRW-spacetime.  Interestingly, in the latter case we discover the appearance of the Heun differential equation, for which the connecting formulas between local solutions have only recently been investigated \cite{Mai07,BILT23}. Their approximation for long time intervals derived in \cite{BILT23} will be used to obtain explicit expression for the Bogolyubov transformations, which then yield the corresponding probabilities for particle creation. Interestingly, for large frequencies, and in the spatially flat case, we find a Planckian frequency spectrum whose temperature is inversely proportional to the duration of scale change. We finish with some concluding remarks in the Outlook, Section~\ref{sect:outlook}.

In the appendix we consider two additional, alternative models for long gravitational pulses with simpler scale functions. They confirm the results obtained in Section \ref{sect:flrw} but are less flexible for a full analysis since both the duration of the time interval and the rate of change of the scale function during the turning on/off are controlled by the same parameter. Instead, the solution to the Heun equation in Section \ref{sect:flrw} allows to take the limit of an infinitely long pulse without reducing the rate of change to become 0.

\subsection*{Acknowledgements}
WvS thanks Erik Koelink for fruitful discussions on the Heun equation.
This work was supported by the ERC Synergy Grant ``BlackHolistic'', the NWO Spinoza Prize, a grant from NWO NWA 6201348, and the Excellence Fellowship from Radboud University. 

\subsection*{Conflict of interest statement} 
The authors declare that there are no conflicts of interest.

\subsection*{Data availability statement} 
Not applicable: no data has been created or analyzed in this study.

\section{The Schwinger effect revisited: a long electric field pulse}
\label{sect:schw}
The derivation of the Schwinger effect is well-documented in the physics literature \cite[Chapter 13]{Sch14}, including the derivation via Bogolyubov transformation as presented here (see \cite{GMM94}). The main reason to include a detailed derivation in the present manuscript is to confront it with particle creation in a gravitational context, as discussed in the subsequent sections. 

\subsection{Quantum scalar field with a background electromagnetic field}
We recall the description of a quantum scalar field in a (classical) electromagnetic background field; see {\em e.g.} \cite{GMM94}. The relevant field equation is
\begin{equation}
  (\partial_\mu + i e A_\mu) (\partial^\mu + i e A^\mu) \phi + m^ 2 \phi =0 .
  \label{eq:eom-A}
  \end{equation}
We consider the four-potential $A_\mu$ to be of the form
$$
A_ \mu(x) = (0,0,0,A_3(t))
$$
so the electromagnetic field is given by
$$
{\bf E}(t) = \left( 0,0,- \frac{dA_3}{dt} \right).
$$
Solutions to \eqref{eq:eom-A} are given by modes of the form
$$
\phi(x) = (2\pi)^{-3/2} (2 \omega_-(\p))^{-1/2} e^ {i \p \cdot \x} \psi_\p(t)
$$
where $\psi_\p(t)$ satisfies the ODE
\begin{equation}
\ddot \psi_\p + \omega^2(\p,t) \psi_\p(t) =0; \qquad \omega^2(\p,t) = m^ 2 + p_\perp^ 2+ (p_3 - eA_3)^2,
\label{eq:ode-A}
\end{equation}
and where we have written $p_\perp^2=p_1^2 + p_2^2$ and $\omega_-(\p) = \lim_{t \to -\infty} \omega(\p,t)$ and have used dots ($\;\dot{}\;$) for derivatives with respect to $t$. 

Introducing the modes for early times
$$
f_\p (x) = (2\pi)^{-3/2} (2 \omega_-(\p))^{-1/2} e^ {i \p \cdot \x} e^{-i \omega_-(\p) t}
$$
we may expand the field as
\begin{equation}
\phi = \int d^ 3 \p \left( A_\p f_\p + A_\p^ \dagger f_\p^* \right)
\label{eq:phi-in}
\end{equation}
with annihilation and creation operators $A_\p,A_\p^\dagger$ at early times.
The strategy is now to look for solutions $\psi_\p$ with positive frequency when $t \to -\infty$ , {\em i.e.}
\begin{equation}
\psi_\p(t) \sim e^{-i \omega_-(\p) t} ; \qquad (t \to -\infty)
\label{eq:psi-p}
\end{equation}
and determine the form of $\psi_\p$ at $t = + \infty$. If we write $\omega_+(\p) = \lim_{t \to \infty} \omega(\p,t)$ then we clearly have 
\begin{equation}
\psi_\p (t) \sim \alpha_\p  e^{-i \omega_+ (\p) t}  + \beta_\p e^ {i \omega_+(\p) t} ; \qquad (t \to\infty),
\label{eq:psi-infty-Schwinger}
\end{equation}
for some coefficients $\alpha_\p, \beta_\p$. In other words, we have a mixture of positive and negative frequency solutions. Since the Wronskian for the ODE \eqref{eq:ode-A} is conserved, we find that
\begin{equation}
  |\alpha_\p|^ 2 - | \beta_\p |^ 2 = 1.
  \label{eq:alphabeta-1-Schwinger}
\end{equation}
Moreover, we may expand the field operator in terms of late time modes
\begin{equation}
\phi = \int d^3 \p \left(a_\p g_\p + a_\p^\dagger g_\p^* \right)
\label{eq:phi-out}
\end{equation}
where now $g_\p \sim (2\pi)^{-3/2} (2 \omega_+(\p))^{-1/2} e^{i \p \cdot x}  e^{-i \omega_+(\p) t}$ (as $t \to \infty$) is the positive frequency late time mode. By comparing Equations \eqref{eq:phi-in} and \eqref{eq:phi-out} we conclude that the annihilation and creation operators $a_\p, a_\p^\dagger$ for the late time quantum field are related to the analogues $A_\p,A_\p^\dagger$ for the early time field via the following {\em Bogolyubov transformation}:
$$
a_\p = \alpha_\p A_\p + \beta_\p^* A_{-\p}^\dagger. 
$$
The expectation value of the number of particles can now be computed as follows. Suppose that the initial state of the system is described in the Heisenberg picture by the vacuum vector $\ket{0}$, {\em i.e.} with no particles at $t \to -\infty$. In other words we have 
$$
A_\p \ket{0}  =0 ; \qquad ( \forall \p).
$$
The expectation value $n(\p)$ of the number of particles present at $t \to \infty$  in mode $\p$ is then given by
$$
n(\p) := \bra{0} a_\p^ \dagger a_\p \ket{0} = | \beta_\p|^ 2,
$$
which yields for the $S$-matrix that 
$$
| \langle 0_{\text{out}} | 0_{\text{in}} \rangle |^2 = \exp \left\{ - V/(2\pi)^3 \int d^3 \p \log (1+n(\p)) \right\}
$$
where the system is placed in a box of volume $V$. Moreover, the probability of observing at late times $n$ particles in mode $\p$  is
$$
P_n(\p ) = \left| \frac{\beta_\p}{\alpha_\p} \right|^{2n} \left (1-  \left| \frac{\beta_\p}{\alpha_\p} \right|^{2 }\right)
$$
so that $n(\p) = \sum_{n \geq 0} n ~ P_n(\p)$.

\subsection{Electric field pulse and hypergeometric equation}
\begin{figure}
\includegraphics[scale=.8]{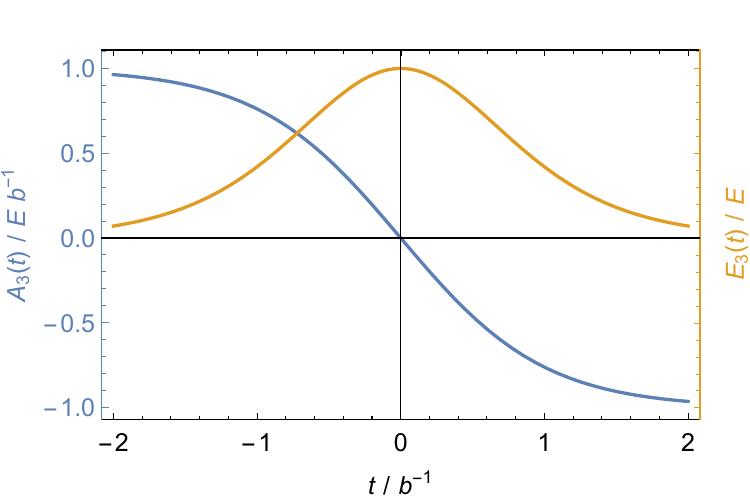}
\caption{Relevant components of vector potential $A_3$ and electric field $E_3$ for an electric field pulse as in Equation~\eqref{eq:AE}.}
\label{fig:AE}
  \end{figure}

We now consider the following explicit vector potential, known as Sauter pulse \cite{Sau32} (see Figure~\ref{fig:AE}):
\begin{equation}
  A_3 (t) = - \frac E{\bp} \tanh \bp t; \qquad {\bf E} (t) =  (0,0,E/ \cosh^2 \bp t) .
  \label{eq:AE}
\end{equation}

Upon writing $u= e^{2 \bp t}$ and denoting derivatives with respect to $u$ by primes ($'$), this becomes $A_3 = -\tfrac E{\bp} \tfrac{u-1}{u+1}$ while \eqref{eq:ode-A} reads
\begin{equation}
\psi_\p''   + u^{-1} \psi_\p' + u^{-2} \left( \left( \frac m{2\bp} \right)^2 + \left( \frac {p_\perp}{2\bp}\right)^2
+ \left( \frac {p_3}{2\bp} + \frac{eE}{2\bp^2} \frac{u-1}{u+1}\right)^2  \right) \psi_\p =0.
\label{eq:ode-A-u}
\end{equation}
This is a ODE with three singular points (at $u=0,-1$ and at $u=\infty$). Let us analyze the behavior at the singular points $u=0,-1$. Around $u=0$ we find that $\psi_\p$ satisfies
$$
\psi_\p'' + u^{-1} \psi_\p' + u^ {-2} \frac{\omega_-^2(\p)}{4 \bp^2} \psi_\p = 0 ,
$$
which is solved by $\psi_\p = u^{\pm c_-}$ where $c_- = i \omega_-(\p)/2\bp$. In order to account for the positive frequency at $u=0$ we take $\psi_\p \sim u^{-c_-}$.

The leading behavior at $u=-1$ is given by
$$
\psi_\p'' + \left( \frac {eE}{\bp^2} \frac {1}{u+1} \right)^ 2 \psi_\p = 0 .
$$
This is solved by $\psi_\p \sim (1+u)^d$ where
$$
d= \tfrac 12 + \sqrt{ \tfrac 14 - \alpha^2}; \qquad \alpha = \frac {eE}{\bp^2}.
$$
If we now write $\psi_\p = u^{-c_-} (1+u)^d F(u)$ we find that $F$ should satisfy the following differential equation:
$$
u(1+u) F'' (u) + \left ( \left(1- i \frac{\omega_-}{\bp} \right) (1+u) + 2d u \right) F' (u) + \left (-\alpha^2 + \frac \alpha{\bp} p_3 + d \left (1- i\frac{\omega_-}{\bp}\right) \right)F(u) = 0 .
$$
This is a hypergeometric differential equation, which is usually written as
$$
z(1-z) \frac {d^2 w}{dz^2} + \left( \chg - (\ahg+\bhg+1)z \right) \frac{dw}{dz} - \ahg \bhg w = 0 .
$$
The local solution at $z =0$ with characteristic exponent $0$ is denoted by ${}_2 F_1 (\ahg,\bhg,\chg;z)$, the {\em hypergeometric function}. In our case of interest we find that the constants $\ahg,\bhg,\chg$ are given by
$$
\ahg= d-i (\omega_-+\omega_+)/2\bp ; \qquad 
\bhg= d-i (\omega_-- \omega_+)/2\bp ; \qquad 
\chg = 1 - i \omega_- /\bp
$$
while $z = -u$ (for an overview, see Table~\ref{table:param-hypergeo}). We conclude that a local solution of \eqref{eq:ode-A-u} at $u=0$ is given by
$$
\psi_\p =  u^{-c_-} (1+u)^d\; {}_2 F_1 \left(d-i \frac{\omega_- + \omega_+}{2 \bp} , d-i \frac{\omega_- - \omega_+}{2 \bp} ,1- i \frac{\omega_-}{\bp} ; -u \right). 
$$
The local behavior of the hypergeometric function at $u=0$ ensures that $\psi_\p \propto u^{-c_-}$ at $u \sim 0$. 

\begin{table}
    \begin{tabular}{cc}
\toprule
Hypergeometric parameters & Pulse parameters \\    
\midrule
$z$ & $-u$ \\
$\ahg$ & $d-i (\omega_-+\omega_+)/2\bp$ \\
$\bhg$ & $d-i (\omega_-- \omega_+)/2\bp$ \\
$\chg$ & $1 - i \omega_- /\bp$ \\
\bottomrule
\end{tabular}

\caption{Relation between the parameters of the electric field pulse and the parameters entering the hypergeometric function. We recall that 
$u=e^{2\bp t}$, 
$d= 1/2 + \left( 1/4 - \alpha^2 \right)^{1/2}$, 
$\alpha = eE/\bp^2$, 
$\omega_\pm(\p) = \lim_{t \to \pm\infty} \omega(\p,t)$.}
\label{table:param-hypergeo}
\end{table}

We may relate the local solution at $u=0$ to the ones at $u=\infty$ since we have \cite[Eq. 15.3.7]{AS64}
\begin{align}
  {}_2 F_1 (\ahg,\bhg,\chg;z) 
  &= \frac{\Gamma(\chg) \Gamma(\bhg-\ahg)}{\Gamma(\bhg) \Gamma(\chg-\ahg)} (-z)^{-\ahg}\; {}_2 F_1 (\ahg,1-\chg+\ahg,1-\bhg+\ahg;1/z) \nonumber \\
  & +   \frac{\Gamma(\chg) \Gamma(\ahg-\bhg)}{\Gamma(\ahg) \Gamma(\chg-\bhg)} (-z)^{-\bhg}\; {}_2 F_1 (\bhg,1-\chg+\bhg,1-\ahg+\bhg;1/z).
  \label{eq:corr-hyperg}
\end{align}
This implies that at $u\sim \infty$ we have
\begin{equation}
\psi_\p \sim  \frac{\Gamma(\chg) \Gamma(\bhg-\ahg)}{\Gamma(\bhg) \Gamma(\chg-\ahg)} u^{-c_-+d-\ahg} + 
 \frac{\Gamma(\chg) \Gamma(\ahg-\bhg)}{\Gamma(\ahg) \Gamma(\chg-\bhg)} u^{-c_- + d - \bhg} 
 = \frac{\Gamma(\chg) \Gamma(\bhg-\ahg)}{\Gamma(\bhg) \Gamma(\chg-\ahg)} u^{c_+} 
 + \frac{\Gamma(\chg) \Gamma(\ahg-\bhg)}{\Gamma(\ahg) \Gamma(\chg-\bhg)} u^{-c_+}
 \label{eq:deriv-hyperg}
 \end{equation}
where we have introduced $c_+ = i \omega_+(\p) /2\bp$. We conclude that the early time annihilation and creation operators are related to the late time ones via Bogolyubov coefficients of the form
$$
\alpha_\p = \frac{\Gamma(\chg) \Gamma(\ahg-\bhg)}{\Gamma(\ahg) \Gamma(\chg-\bhg)} \left( \frac{\omega_+(\p)}{\omega_-(\p)}\right)^{1/2}; \qquad
\beta_\p = \frac{\Gamma(\chg) \Gamma(\bhg-\ahg)}{\Gamma(\bhg) \Gamma(\chg-\ahg)} \left( \frac{\omega_+(\p)}{\omega_-(\p)}\right)^{1/2}.
$$
We use the identities $\Gamma(1-z) \Gamma(z) =\pi/ \sin (\pi z)$,  $\overline {\Gamma(z)} = \Gamma(\overline z)$ as well as $\sin(x-y)\sin(x+y) = \sin^2 x - \sin^2 y$ to derive 
$$
\left| \frac {\beta_\p }{\alpha_\p} \right|^2 
= \frac{ \sin^2 \pi d + \sinh^2 ( {\pi (\omega_-(\p)- \omega_+(\p))}/{2\bp}) }{ \sin^2 \pi d + \sinh^2 ( {\pi (\omega_- (\p)+\omega_+(\p))}/{2\bp}) }
$$
while for the expectation value $n(\p)$ we find
$$
n(\p) = | \beta_\p|^2 
=  \frac{ \sin^2 \pi d + \sinh^2 (\pi (\omega_-(\p)- \omega_+(\p))/2\bp) }{ \sinh (\pi  \omega_-(\p)/\bp) \sinh (\pi \omega_+(\p)/\bp)}.
$$

\subsection{Particle creation for a long electric field pulse}
We will now analyze these coefficients for small $\bp$, that is, of a long electric field pulse. 

First, note that
$$
\omega_\pm(\p)^ 2 
= \lim_{t \to \pm \infty} \left( m^2+ p_\perp^2  + \left(p_3- \frac{eE}{\bp} \tanh \bp t \right)^2 \right) 
= m^2 + p_\perp^2 + \left(p_3 \mp \frac {eE}{\bp}\right)^2.
$$
From this we see that $n(\p)$ is exponentially suppressed whenever $|p_3| \gg eE/\bp$.
Moreover, upon expanding the square root in powers of $1/\bp$:
$$
\omega_\pm(\p) = \frac{eE}{\bp} \mp p_3+ \frac{m^2+ p_\perp^2}{2eE} \bp + \O(\bp^2),
$$
while
$$
d = \tfrac 12 + \sqrt{ \tfrac 14 - \alpha^2} = i \frac{eE}{\bp^2} + \O(1).
$$
Using these expressions, we find that for $|p_3| \leq eE/\bp$ ({\em cf.} \cite[Eq. 4.80]{GMM94})
\begin{align*}
  n(\p)& \sim \frac{ \sinh^2 \pi \frac{eE}{\bp^2} + \sinh^2(\pi p_3/\bp) }{ \sinh \pi  \left( \frac{eE}{\bp^ 2} -\frac{p_3}{\bp} + \frac{m^2+ p_\perp^2}{2eE} \right)  \sinh \pi  \left( \frac{eE}{\bp^2} +\frac{p_3}{\bp} + \frac{m^2+ p_\perp^2}{2eE} \right) }
  \\
  &\sim
\frac{ \exp\left\{ 2 \pi \frac{eE}{\bp^2} \right\} }{\exp \left\{ \pi  \left(\frac{2 eE}{\bp^2}  + \frac{m^2+ p_\perp^2}{eE} \right) \right \} }= \exp\{ -\pi (m^2+p_\perp^2)/eE \} ; \qquad (\bp \to 0).
\end{align*}
We thus find that for $\bp \to 0$:
$$
n(\p) \sim \exp\{ -\pi (m^2+p_\perp^2)/eE \} \theta( eE/\bp - |p_3|),
$$
in terms of the Heaviside $\theta$-function.  We use this to obtain 
$$
| \langle 0_{\text{out}} | 0_{\text{in}} \rangle |^2 = \exp \left\{ - V/(2\pi)^3 \int d^3 \p \log (1+n(\p)) \right\}.
$$
When expanding the logarithm and performing the integration over $\mathbf p$ we obtain Schwinger's result \cite{Sch51} ({\em cf.} \cite[Eq. 4.83]{GMM94}):
$$
\frac{1}{(2\pi)^3}\int d^3 \p \log (1+n(\p)) = \frac{e^2 E^2T} {8 \pi^3} \sum_{n=1}^\infty \frac{(-1)^n}{n^2} \exp\left\{ - \frac{n \pi  m^2} {eE} \right\}
$$
where $T =1/(2\bp)$.
For the probabilities $P_n(\p)$ we first determine that
$$
\left| \frac {\beta_\p }{\alpha_\p} \right|^2 = \frac{1}{1+ \exp \left\{ \pi (m^2+p_\perp^2)/e E \right \} }.
$$
For large momenta we have $\left|  {\beta_\p }/{\alpha_\p} \right|^2  \sim \exp \{- \pi (m^2+p_\perp^2)/e E \}$ and, accordingly,
$$
P_n(\p) =  \exp\{- n \pi (m^2+p_\perp^2)/e E \} \left( 1-  \exp\{- \pi (m^2+p_\perp^2)/e E \} \right).
$$

\section{Particle creation in FLRW-spacetimes with a gravitational pulse}
\label{sect:flrw}
\subsection{Scalar fields in FLRW-spacetime}
We briefly recall the description of a free hermitian quantum scalar field in an FLRW-spacetime background, referring to {\em e.g.} \cite[Chapter 2]{PT09} for more details. We write the metric on $\R \times \Sigma$ as
$$
ds^2 = -dt^2 + \a(t)^2 d s_\Sigma^2
$$
for some scale function $\a(t)$, and where $d s_\Sigma^2$ is a fixed volume form on a three-dimensional Riemannian manifold $\Sigma$. The free field equation $\Box \phi =0$ becomes
\begin{equation}
  \a^{-3} \partial_t (\a^3 \partial_t \phi) + \a^{-2} \Delta_\Sigma \phi =0,
  \label{eq:eom}
  \end{equation}
in terms of the spatial Laplacian $\Delta_\Sigma$. The latter has positive eigenvalues: $\Delta_\Sigma \phi_\lambda = \lambda^2 \phi_\lambda$ with $\lambda^2 \in \sigma(\Delta_\Sigma) \subseteq\R$, the spectrum of $\Delta_\Sigma$; we also write $\phi_{-\lambda} = \phi_\lambda^*$.\footnote{The reader should be warned that we adopt the mathematicians' convention of writing $\lambda$ for an eigenvalue of the Laplacian, in contrast to the physicists' convention of writing $\lambda$ for a wavelength. As a subscript, $\lambda$ refers to a unique mode irrespective of potential multiplicities.} In terms of a time variable $\tau$ defined by
\begin{equation}
\tau = \int^t \a^{-3}(t') dt'
\label{eq:tau}
\end{equation}
the field equation \eqref{eq:eom} becomes
\begin{equation}
\frac{\partial^2 \phi}{  \partial \tau^2} + \a^4(\tau) \Delta_\Sigma \phi =0.
  \label{eq:eom-tau}
\end{equation}
If $\phi_\lambda$ are the eigenfunctions $\phi_\lambda$ of $\Delta_\Sigma$, we may decompose the field operator $\phi$ as
\begin{equation}
\phi = \sum_\lambda A_\lambda f_\lambda + A_\lambda^\dagger f_\lambda^*
\end{equation}
in terms of respective annihilation and creation operators $A_\lambda, A_\lambda^\dagger$ and where $f_\lambda =  \phi_\lambda \psi_\lambda $. The function $\psi_\lambda$ is a function of $\tau$ only and from \eqref{eq:eom} we derive that it satisfies the following ODE
\begin{equation}
\frac{d^2 \psi_\lambda}{  d \tau^2} + \lambda^2 \a^4(\tau)\psi_\lambda =0.
\label{eq:ode-tau}
\end{equation}
We consider an asymptotically static $\a(\tau)$ and impose the initial condition that at $\tau \to -\infty$ the function $f_\lambda$ is the positive frequency solution (at scale $a_{-\infty} = \lim_{\tau \to -\infty} \a(\tau)$), {\em i.e.}
\begin{equation}
  f_\lambda \sim \phi_\lambda e^ {-i \omega_\lambda t} \equiv  \phi_\lambda e^{-i \omega_\lambda a_{-\infty}^ 3 \tau}  ; \qquad (\tau \to -\infty),
  \label{eq:normalmode}
\end{equation}
where $\omega_\lambda = | \lambda|/a_{-\infty}$.
The strategy is now to look for solutions $\psi_\lambda$ satisfying this initial condition, and determine the form of $\psi_\lambda$ at $\tau = + \infty$. If we write $a_\infty = \lim_{\tau \to \infty} \a(\tau)$ then we clearly have for $\tau \to \infty$:
\begin{equation}
\psi_\lambda (\tau) \sim \alpha_\lambda  e^{-i |\lambda| a_\infty^2 \tau} + \beta_\lambda e^ {i |\lambda| a_\infty^2 \tau} ; \qquad (\tau \to\infty),
\label{eq:psi-infty-gravitational}
\end{equation}
for some coefficients $\alpha_\lambda, \beta_\lambda$. In other words, we have a mixture of positive and negative frequency solutions; as we will see, and as is well-known, this is the origin of particle production in a curved spacetime. Since the Wronskian for the ODE \eqref{eq:ode-tau} is conserved, we find that
\begin{equation}
  |\alpha_\lambda|^ 2 - | \beta_\lambda |^ 2 = 1.
  \label{eq:alphabeta-1-gravitational}
\end{equation}
Moreover, we may expand the field operator in terms of late time modes
$$
\phi = \sum_\lambda a_\lambda g_\lambda + a_\lambda^\dagger g_\lambda^*
$$
where now $g_\lambda \sim \phi_\lambda  e^{-i |\lambda| a_\infty^2 \tau}$ is the positive frequency late time mode. We conclude that the creation and annihilation operator $a_\lambda, a_\lambda^\dagger$ for the late time quantum field are related to the analogues $A_\lambda, A_\lambda^\dagger$ for the early time field via the following Bogolyubov transformation:
$$
a_\lambda = \alpha_\lambda A_\lambda + \beta_\lambda^* A_{-\lambda}^\dagger. 
$$
The expectation value of the number of particles can now be computed as follows. Suppose that the initial state of the system is described in the Heisenberg picture by the vacuum vector $\ket{0}$, {\em i.e.} with no particles at $\tau \to -\infty$. In other words we have 
$$
A_\lambda \ket{0}  =0 ; \qquad ( \forall \lambda \in \sigma(\Delta_\Sigma)).
$$
The expectation value $n(\lambda)$ of the number of particles present at $\tau \to \infty$  in mode $\lambda$ is then given by
\begin{equation}
  n(\lambda) = \bra{0} a_\lambda^ \dagger a_\lambda \ket{0} = | \beta_\lambda|^ 2 = \frac{|\beta_\lambda/\alpha_\lambda     |^2}{1-|\beta_\lambda/\alpha_\lambda|^2},
  \label{eq:n-lambda}
\end{equation}
which yields for the $S$-matrix that
$$
| \langle 0_{\text{out}} | 0_{\text{in}} \rangle |^2 = \exp \left\{ - \sum_\lambda \log (1+n(\lambda) ) \right\}.
$$
Moreover, the probability of observing at late times $n$ particles in mode $\lambda$  is
\begin{equation}
  P_n(\lambda ) = \left| \frac{\beta_\lambda}{\alpha_\lambda} \right|^{2n} \left (1-  \left| \frac{\beta_\lambda}{\alpha_\lambda} \right|^{2 }\right).
  \label{eq:prob-n}
\end{equation}

\subsection{Particle creation in instantaneous long gravitational pulse: Heun equation}
\label{sect:heun}
\begin{figure}
  \subfigure[Scale function $\a(\tau)$ for $T=15 b^{-1}$.]{\includegraphics[scale=.48]{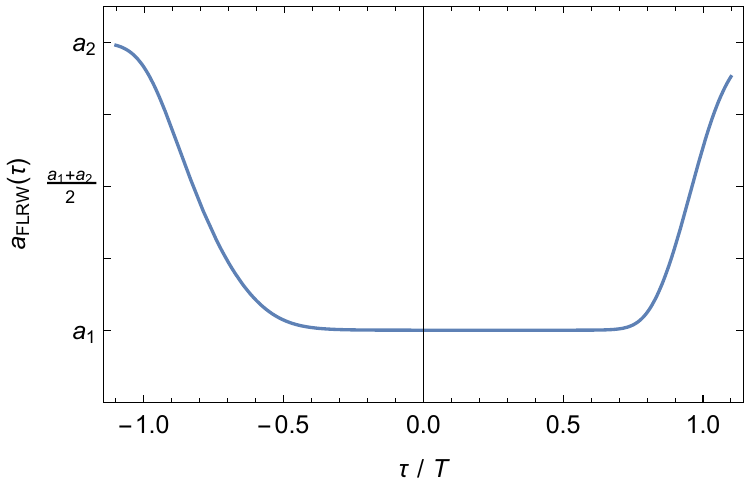}}
  \hspace{5mm}
  \subfigure[Scale function $\a(\tau)$ for $T=50 b^{-1}$.]{\includegraphics[scale=.48]{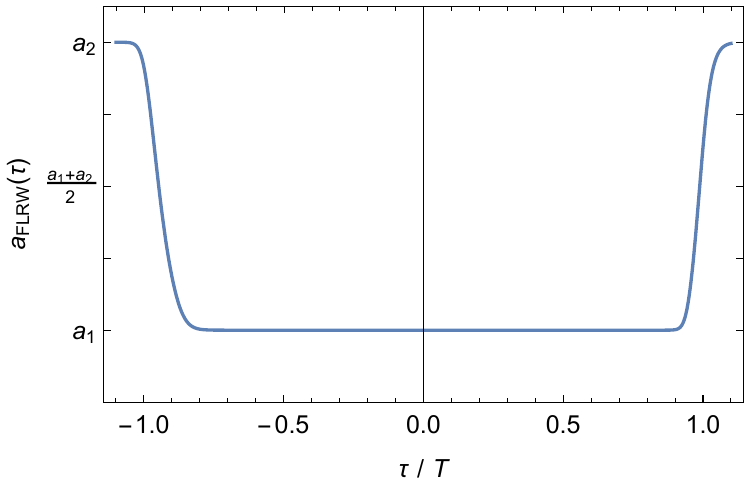}}\\
  \caption{Scale function $\a(\tau)$ as in Eq. \eqref{eq:scale-FLRW} ($a_1=1, a_2=5$).}
  \label{fig:a}
\end{figure}

\begin{figure}  
  \subfigure[Scalar curvature for spatial curvature $\kappa=0$.]{\includegraphics[scale=.48]{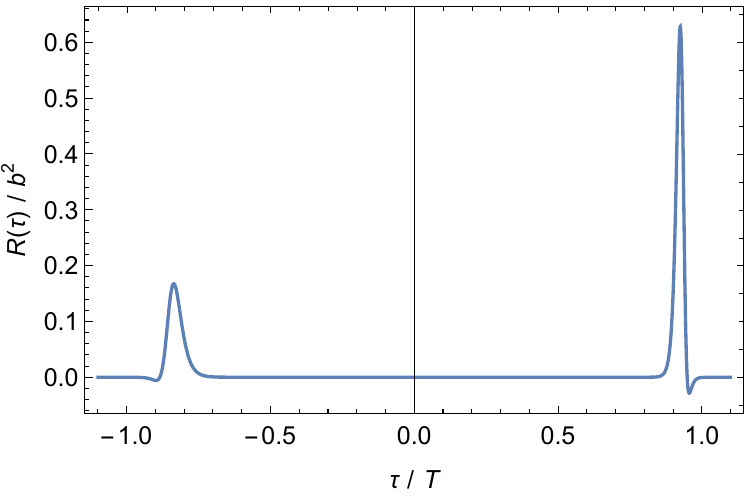}}\hspace{5mm}
  \subfigure[Scalar curvature for spatial curvature $\kappa=1$.]{\includegraphics[scale=.48]{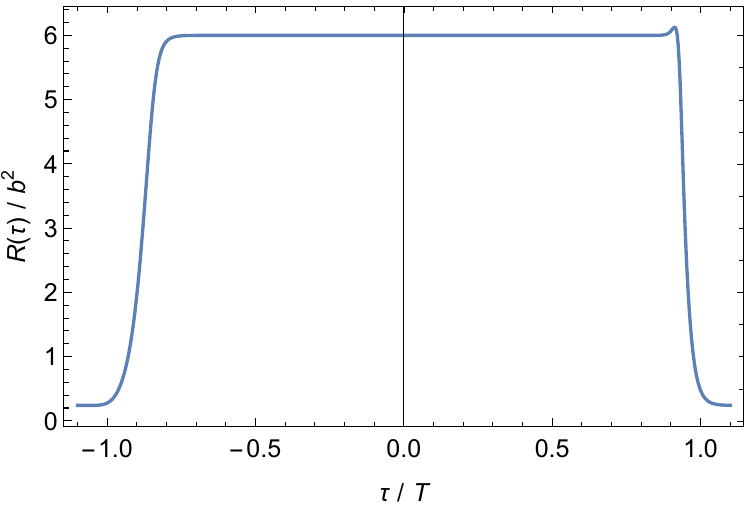}}\\
  \subfigure[Scalar curvature for spatial curvature $\kappa=-1$.]{\includegraphics[scale=.48]{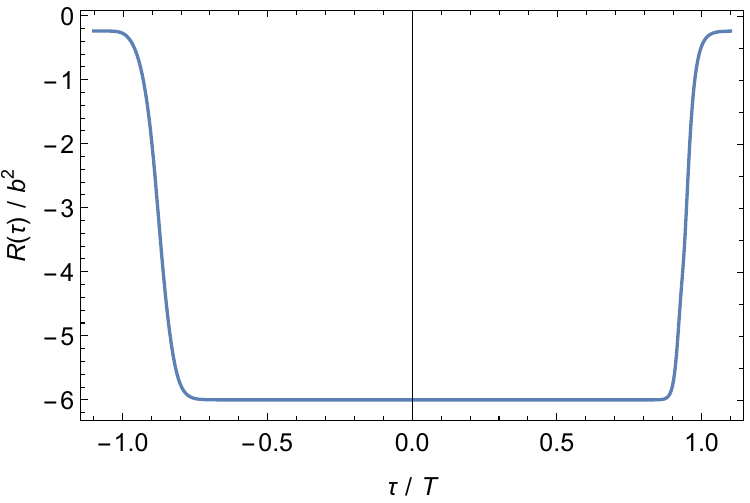}}
\caption{Scalar curvature $R(\tau)$ for FLRW-spacetime with a gravitational pulse as in Eq. \eqref{eq:scale-FLRW} and \eqref{eq:ricci}, 
displayed at same scaling ($a_1=1$, $a_2=5$, $T=50 b^{-1}$).}
\label{fig:a-curv}
  \end{figure}

The above ODE \eqref{eq:ode-tau} can be solved exactly in some cases, see {\em e.g.} \cite[Section 2.8]{PT09} and references therein. Here we will analyze an FLRW-model for a pulse in the scalar function $\a(\tau)$ where we include two parameters to control the shape: a parameter $T$ for the time interval of the pulse, and a slope parameter $b$ for the rate of change of the scale function when the pulse is `turned on/off'. More precisely, we take 
\begin{equation}
  \a(\tau) = \left( a_2^4 + \left( a_2^4- a_1^4\right) \left (  \left(\frac {e^{b(\tau-T)}}{1+ e^{b(\tau-T)}} \right)^2 -   \left(\frac {e^{b(\tau+T)}}{1+ e^{b(\tau+T)}} \right)^2 \right) \right)^{1/4}.
  \label{eq:scale-FLRW}
\end{equation}
Recall that the Ricci scalar curvature for FLRW spacetimes is given in the time variable $t$ by
$$
R(t) =  6  \left( \frac {\addot(t)}{\a(t)} +  \frac {\adot(t)^2}{\a(t)^2} + \frac {\kappa}{\a(t)^2} \right)
$$
where $\kappa$ is the Gaussian curvature for the spatial geometry $\Sigma$. In terms of the parameter $\tau$ one finds that
\begin{align}
  R(\tau) = 6 \left( \frac{\a''  (\tau)}{\a(\tau)^{7}}  - 2\frac{\a' (\tau)^2}{\a(\tau)^{8}} +\frac{\kappa}{ \a(\tau)^{2}} \right).
  \label{eq:ricci}
\end{align}
We have illustrated the scale function in Figure \ref{fig:a} and the corresponding curvature scalars in Figure \ref{fig:a-curv}.

A useful coordinate transformation is $u = e^{b\tau}$ while setting $u_\pm = e^{\pm b T}$. Then \eqref{eq:eom-tau} becomes for the above $\a(\tau)$:
\begin{equation}
  \psi_\lambda '' + u^{-1} \psi_\lambda' + \lambda^2 b^{-2} \left( u^{-2} a_2^4 + \left( \frac 1 {(u_+ + u)^2} - \frac 1 {(u_- + u)^ 2} \right) \left ( a_2^ 4-a_1^4 \right) \right) \psi_\lambda = 0 ,
  \label{eq:eom-u}
\end{equation}
where the prime denotes a derivative with respect to $u$.

Let us analyze the behavior of the differential equation \eqref{eq:eom-u} at the singular points $u=0, u=-u_+$, and $u=-u_-$.
At $u =0$ ({\em i.e.} $\tau \to -\infty$) the leading behavior is then
$$
\psi_\lambda'' + u^{-1} \psi_\lambda ' +  \lambda^2 b^{-2}  u^{-2} a_2^4 \psi_\lambda = 0; \qquad ( u \sim 0)
$$
with solutions near $u=0$ given by $\psi_\lambda \sim u^{\pm c}$ where $c =i |\lambda|b^{-1} a_2^ 2$. In order to account for the positive frequency condition at $u = 0$ we take $\psi_\lambda \sim u^{-c}$. 

At $u=-u_\pm$ we find leading behavior
$$
\psi_\lambda''  \pm  \lambda^2 b^{-2} (a_2^4 - a_1^4) \frac 1 {(u_\pm + u)^2}  \psi_\lambda = 0; \qquad ( u \sim -u_\pm)
$$
with solutions given by $\psi_\lambda \sim (u_\pm + u)^{d_\pm}$ where
$$
d_\pm (d_\pm -1 ) = \mp \lambda^ 2b^{-2} (a_2^4 - a_1^4) .
$$
This suggests the following Ansatz for solutions of \eqref{eq:eom-u} of the form
$$
\psi_\lambda (u) = u^{-c} (u_-+u)^{d_-} (u_++u)^{d_+} h(u)
$$
where then $h(u)$ is found to satisfy the following differential equation:
\begin{equation}
  h'' (u) +   \left( \frac \gamma u + \frac {2d_+} {u+u_+} + \frac {2d_-} {u+u_-} \right) h' (u) + \frac{(\gamma(d_+ + d_-) + 2d_+ d_- ) u + \gamma (d_+ u_- + d_- u_+ )}{u(u+u_-)(u+u_+) } h(u) = 0
  \label{eq:ode-h}
\end{equation}
in terms of  $\gamma = 1- 2c$.

The above differential equation can be recognized as the Heun differential equation. Indeed, Heun considered \cite{Heu1889} ({\em cf.} \cite{Ron95} and references therein) the following differential equation with four singular points (at $z=0,a,1$ and $\infty$):
\begin{equation}
  \frac {d^ 2 y } {dz^ 2} + \left ( \frac \gamma z + \frac \delta {z-1} + \frac{\epsilon}{ z-a} \right) \frac {dy }{dz}+ \frac {\alpha \beta z - q}{z(z-1)(z-a)}  y = 0,
  \label{eq:heun}
\end{equation}
where the coefficients satisfy the Fuchsian condition $\gamma + \delta + \epsilon = \alpha + \beta + 1$.\footnote{The constants $\alpha,\beta$ in the Heun equation should not be confused with the Bogolyubov coefficients $\alpha_\lambda,\beta_\lambda$ which in fact always carry a subscript.} Let us denote the local solution of \eqref{eq:heun} near $z= 0$ for which $y(z) \sim 1 + \O(z)$ by $H\ell(a,q;\alpha,\beta, \gamma, \delta; z)$. 

Returning to our case of interest, \eqref{eq:ode-h} is indeed a Heun differential equation with constants related to our parameters $\gamma, u_\pm, c, d_\pm $ by
$$
\gamma=\gamma,\qquad  a = \frac{u_-}{u_+}, \qquad  \delta= 2d_+, \qquad \epsilon = 2d_-,
$$
and for which, after including also the Fuchsian condition, we find
\begin{equation}
\alpha = -2c + d_+ + d_- , \qquad \beta = d_+ + d_-. 
\label{eq:const-Heun}
\end{equation}
We conclude that a local solution of \eqref{eq:eom-u} at $u =0$ is given (for these values of the constants, see also Table \ref{table:param}) by 
$$
\psi_\lambda(u) = u^{-c} (u_-+u)^{d_-} (u_++u)^{d_+} H\ell \left(a, q;\alpha,\beta, \gamma, \delta; - u /{u_+} \right).
$$
One may check that the leading behavior of the local Heun function $H\ell$ ensures that $\psi_\lambda \propto u^{-c}$ at $u \sim 0$, as desired.

In order to analyze the behavior of the large time solutions, and to determine the Bogolyubov coefficients, we will use the relations between local solutions of the Heun equations around different singular points as derived in \cite{Mai07}. 
We are mainly interested in expressing the function 
$H\ell \left(a, q;\alpha,\beta, \gamma, \delta; - u /{u_+} \right)$ as a linear combination of a pair of local solutions around $u = +\infty$, in such a way that the behavior of $\psi_\lambda$ is as indicated in \eqref{eq:psi-infty-gravitational}.

From \cite[Table 2]{Mai07} we conclude that $H\ell \left(a, q;\alpha,\beta, \gamma, \delta; - u /{u_+} \right)$ can be written as a linear combination of two local solutions at $\infty$ with characteristic exponents $\alpha$ and $\beta$, respectively: 
\begin{multline}
H\ell \left(a, q;\alpha,\beta, \gamma, \delta; x \right) = \\
  C_1 x^{-\alpha} H\ell \left(a, q- \alpha \beta (1+a) + \alpha (\delta + a \epsilon) ;\alpha, \alpha-\gamma +1, \alpha -\beta+1, \alpha + \beta - \gamma -\delta+1; \tfrac a x\right) \\
    +C_2 x^{-\beta} H\ell \left(a, q- \alpha \beta (1+a) + \beta (\delta + a \epsilon) ;\beta, \beta-\gamma +1, -\alpha + \beta +1, \alpha + \beta - \gamma -\delta +1; \tfrac  a x\right)
  \label{eq:connecting}
\end{multline}
for some constants $C_1,C_2$.\footnote{In \cite[Table 2]{Mai07} these local solutions can be found to be labeled by the Coxeter group elements $[\infty_+0_+][1_+a_+]$ and $[\infty_-0_-][1_+a_+]$, respectively.} 
We can then read off the behavior of the solution $\psi_\lambda$ at $u=\infty$ by considering these two local solutions; and find
$$
\psi_\lambda(u) \sim C_1 u^{-c+d_+ + d_- -\alpha } + C_2 u^{-c+d_+ +d_- -\beta} = C_1 u^{c} +  C_2 u^{-c},
$$
using the values of $\alpha$ and $\beta$ in Equation \eqref{eq:const-Heun}. Since $c= i |\lambda|b^{-1} a_2^2$, we conclude that the Bogolyubov coefficients in Equation \eqref{eq:psi-infty-gravitational} are related to the two constants $C_1,C_2$ via
$$
\alpha_\lambda = C_2; \qquad \beta_\lambda = C_1.
$$
In the next section, we will explicitly determine these coefficients in the limit of a large time interval, that is, when $a = u_-/u_+ =e^{-2b T} \to 0$. This then allows to compute the probability for particle production.

\subsection{Particle creation for a sudden, long gravitational pulse}
Connecting formulas for local Heun functions around different singular points (such as \eqref{eq:connecting} have been derived as semi-classical limits of connecting formulas for conformal blocks by the authors of \cite{BILT23}. In fact, they translate the Heun equation to a semi-classical BPZ equation satisfied by certain five-point CFT-correlation functions, exploiting a dictionary between the parameters involved. Then, by expanding these solutions to the BPZ equation in two different regions, and making use of crossing symmetry of the CFT correlation functions, they arrive at the connecting formulas. In turn, these translate to the connecting formulas for the Heun functions, using the dictionary in Table \ref{table:param}. 

Concretely, these connection formulas yields the desired expressions for the Bogolyubov coefficients $\alpha_\lambda$ and $\beta_\lambda$ as a power series expansion in $a= u_-/u_+ = e^{-2bT}$.\footnote{Note that our parameter $a$ corresponds to the parameter $t$ in \cite{BILT23}, which we reserve for time in the FLRW-spacetime.} Since we are interested in a long gravitational pulse, {\em i.e.} $T \to \infty$, we will limit ourselves to the leading behavior up to $\O(a)$. Then, \cite[Eq. (4.1.22)]{BILT23} yields
\begin{align*}
\alpha_\lambda &= \sum_{\sigma = \pm} \frac{\Gamma(1-2 \sigma a(q)) \Gamma(-2 \sigma a(q) ) \Gamma(\gamma) \Gamma(\alpha-\beta)  a^{\frac{\gamma+\epsilon-1}{2}-\sigma a(q)} e^ {i \pi (\delta+\gamma)/2} }
      {\Gamma( \frac{\gamma-\epsilon+1}{2}-\sigma a(q) ) \Gamma (\frac{\gamma+\epsilon-1}{2}-\sigma a(q) ) \Gamma( 1+ \frac{\alpha-\beta-\delta}{2}-\sigma a(q) )\Gamma ( \frac{\alpha-\beta+\delta}{2}-\sigma a(q) ) } + \O(a)
      \intertext{and}
\beta_\lambda &= \sum_{\sigma = \pm} \frac{\Gamma(1-2 \sigma a(q)) \Gamma(-2 \sigma a(q) ) \Gamma(\gamma) \Gamma(\beta-\alpha)  a^{\frac{\gamma+\epsilon-1}{2}-\sigma a(q)} e^ {i \pi (\delta+\gamma)/2} }
      {\Gamma( \frac{\gamma-\epsilon+1}{2}-\sigma a(q) ) \Gamma (\frac{\gamma+\epsilon-1}{2}-\sigma a(q) ) \Gamma( 1+ \frac{\beta-\alpha-\delta}{2}-\sigma a(q)) \Gamma ( \frac{\beta-\alpha+\delta}{2}-\sigma a(q) ) }  + \O(a)
\end{align*}
and where $a(q)$ is given by
$$
a(q) = \frac 12 \sqrt{3-4q + \gamma^ 2 + 2 \gamma(\epsilon-1) + \epsilon(\epsilon-2) } +\O(a).
$$
This expression is obtained by inverting a combinatorial identity that is satisfied by the parameter $q$ in terms of the momentum parameter; again, it is derived in the semi-classical limit (see \cite[Appendix C]{BILT23}). 
\begin{table}
    \begin{tabular}{cc}
\toprule
Heun parameters & Gravitational parameters \\    
\midrule
$z$ & $-u/u_-$ \\
$a$ & $u_-/u_+= e^{-2bT}$ \\
$q$ & $\gamma/u_+ (d_+ u_- + d_- u_+)$ \\
$\alpha$&  $-2c+d_++ d_-$ \\
$\beta$ & $d_++ d_-$  \\
$\gamma$ &$\gamma =1-2c$ \\
$\delta$ &$2d_+$ \\
$\epsilon$ & $2d_-$ \\
\bottomrule
\end{tabular}
\caption{Relation between the gravitational parameters and the  parameters entering the Heun function. Note that $a$ is denoted by $t$ in \cite{BILT23}, which we reserve for time in the FLRW-spacetime. Moreover, we recall that $u=e^{b\tau}, u_\pm = e^{\pm bT }$, $c=i |\lambda| b^{-1} a_2^2$ and $d_\pm (d_\pm -1 ) = \mp \lambda^ 2b^{-2} (a_2^4 - a_1^4)$.}
\label{table:param}
\end{table}

Using Table \ref{table:param} once again, we can express this in terms of the parameters $c,d_1,d_2$ as well, finding that $a(q) = \sqrt{\tfrac 12- \lambda^2 b^{-2} a_1^4} +\O(a)$ and
\begin{align*}
  \alpha_\lambda &= \left( A_{+}(\lambda)  a^{-c+d_--a(q)} + A_{-}(\lambda)  a^{-c+d_-+a(q)} \right) (1 +\O(a));\\
  \beta_\lambda &= \left(B_{+}(\lambda)  a^{-c+d_--a(q)} + B_{-}(\lambda)  a^{-c+d_- + a(q)} \right)(1 +\O(a)),
  \end{align*}
where for $\sigma= \pm$ we have
     \begin{align*}
       A_{\sigma}(\lambda) &= \frac{\Gamma(1-2 \sigma a(q)) \Gamma(-2 \sigma a(q) ) \Gamma(1-2c) \Gamma(-2c) e^ {i \pi (\delta+\gamma)/2} }
{\Gamma( 1-c-d_--\sigma a(q) ) \Gamma (-c+d_--\sigma a(q) ) \Gamma( 1-c-d_+ -\sigma a(q)) \Gamma ( -c+d_+-\sigma a(q) ) },
\intertext{and}
B_\sigma(\lambda) &= \frac{\Gamma(1-2 \sigma a(q)) \Gamma(-2 \sigma a(q) ) \Gamma(1-2c) \Gamma(2c)   e^ {i \pi (\delta+\gamma)/2} }
    {\Gamma( 1-c-d_--\sigma a(q) ) \Gamma (-c+d_--\sigma a(q) ) \Gamma( 1+c-d_+-\sigma a(q) )\Gamma ( c+d_+-\sigma a(q) ) }.
\end{align*}
We identify three different regions for $|\lambda|$ for which the constants $a(q), d_\pm$ are real or contain an imaginary part. Indeed, we have
$$
d_\pm = \frac 1 2 \left( 1+ \sqrt{ 1\mp 4 \lambda^2 b^{-2} (a_2^4-a_1^4)} \right); \qquad a(q) = \sqrt{\tfrac 12- \lambda^2 b^{-2} a_1^4}
$$
so that when we assume that $a_2^4 > 3 a_1^4/2$ we may distinguish three intervals $[0,\lambda_m], [\lambda_m,\lambda_M]$ and $[\lambda_M,\infty)$ where
  $$
\lambda_m^2 = \frac{b ^2}{4(a_2^4-a_1^4)}; \qquad \lambda_M^2 = \frac{b^2}{2a_1^4}.
  $$
The resulting behavior of the coefficients $a(q)$ and $d_+$ are then:
\begin{center}
\begin{tabular}{c@{\qquad}c@{\qquad}c@{\qquad}c}
\toprule
&$|\lambda| \leq \lambda_m $
  &$ \lambda_m \leq |\lambda| \leq \lambda_M$ & $\lambda_M \leq |\lambda|$ \\
\midrule
  $a(q) $ & $\sqrt{\tfrac 12- \lambda^2 b^{-2} a_1^4}$ & $  \sqrt{\tfrac 12- \lambda^2 b^{-2} a_1^4}$& $i  \sqrt{-\tfrac 12+ \lambda^2 b^{-2} a_1^4}$\\
  $    d_+ $& $\frac 1 2 \left( 1+ \sqrt{ 1-4 \lambda^2 \frac{a_2^4-a_1^ 4} {b^ 2} } \right)$
  & $\frac 1 2 \left( 1+ i \sqrt{ -1+ 4 \lambda^2 \frac{a_2^4-a_1^ 4} {b^ 2}} \right)$
  & $\frac 1 2 \left( 1+ i \sqrt{ -1+4 \lambda^2 \frac{a_2^4-a_1^ 4} {b^ 2} } \right)$\\
\bottomrule
\end{tabular}
\end{center}
where all the square roots that appear in this table are real-valued and $d_-$ is positive for all values of $\lambda$.

Let us analyze the profile for potential particle production in these three eigenvalue regions. Recall that for late times the eigenvalue $\lambda$ is related to the frequency $\omega_\lambda$ of a late time normal mode via $\omega_\lambda = |\lambda| / a_2$ ({\em cf.} Equation \eqref{eq:normalmode}).
 
\subsubsection{Low frequency modes}
If $|\lambda| \leq \lambda_M$ we are in the low frequency range, since $\omega_\lambda \leq \lambda_M/a_2 = b/ (\sqrt{2} a_1^2 a_2)$. 
We will compute the ratio $|\beta_\lambda|^2/ |\alpha_\lambda|^2$ in this range.
Since $a(q) \geq 0$ we find that 
$$
\alpha_\lambda = A_+(\lambda) a^{-c+d_-- a(q)} +\O(a); \qquad \beta_\lambda = B_+ (\lambda) a^{-c+d_-- a(q)} +\O(a)
$$
and, consequently, in the limit of large time interval:
$$
\lim_{a \to 0} \frac{\beta_\lambda}{\alpha_\lambda} =  \frac{B_+(\lambda)}{A_+(\lambda)}
= \frac{ \Gamma(2c)}{\Gamma(-2c)}\frac{\Gamma(1-c-d_+ -a(q)) \Gamma(-c+d_+ - a(q) )}{\Gamma(1+c-d_+-a(q)) \Gamma(c+d_+ -a(q))} .
$$
We now distinguish two cases: $|\lambda| \leq \lambda_m$ and $\lambda_m \leq |\lambda| \leq  \lambda_M$. In the first region, $d_+$ is real-valued, so that, using $\overline{\Gamma(z)} = \Gamma(\overline{z})$ we find with $c \in i \R$ that
\begin{align*}
\lim_{a \to 0} \frac{|\beta_\lambda|^2}{|\alpha_\lambda|^2} &=
\frac{\Gamma(2c)\Gamma(-2c)}{\Gamma(-2c)\Gamma(2c) }\\ & \quad \times \frac{\Gamma(1-c-d_+ -a(q))\Gamma(1+c-d_+ -a(q)) \Gamma(-c+d_+ - a(q) ) \Gamma(c+d_+ - a(q) )}{\Gamma(1+c-d_+-a(q))  \Gamma(1-c-d_+-a(q)) \Gamma(c+d_+ -a(q)) \Gamma(-c+d_+ -a(q))} =1.
\end{align*}
In the second region, $d_+ \in \frac 12 + i \R$ so that $\overline{d_+} = 1-d_+$. This implies that 
\begin{align*}
\lim_{a \to 0} \frac{|\beta_\lambda|^2}{|\alpha_\lambda|^2} &=
\frac{\Gamma(2c)\Gamma(-2c)}{\Gamma(-2c)\Gamma(2c)} \\ & \quad \times \frac{\Gamma(1-c-d_+ -a(q))\Gamma(c+d_+ -a(q)) \Gamma(-c+d_+ - a(q) ) \Gamma(1+c-d_+ - a(q) )}{\Gamma(1+c-d_+-a(q))  \Gamma(-c+d_+-a(q)) \Gamma(c+d_+ -a(q)) \Gamma(1-c-d_+ -a(q))}=1.
\end{align*}
We conclude that for all $|\lambda| \leq \lambda_M$ we have that $\lim_{a \to 0} \frac{|\beta_\lambda|^2}{|\alpha_\lambda|^2} =1$ so that no finite particle production appears for the low frequencies in the limit of a long time interval.

\subsubsection{High frequency modes}
We now let $|\lambda| \geq \lambda_M$, so that besides $c$ and $d_+-1/2$, now also $a(q)$ is purely imaginary, at least up to $\O(a)$. Let us write
\begin{equation}
  c=i \rho , \qquad d_+ = \frac 12 + i \mu_+ , \qquad a(q) = i \nu,
  \label{eq:imagin-var}
\end{equation}
while we set $d_- = \frac 12 + \mu_-$ (real). Explicitly, we have
\begin{equation}
\rho =  |\lambda| b^{-1} a_2^2; \qquad 
\mu_\pm = \sqrt{\mp \frac 14 +  \lambda^2 b^{-2} (a_2^ 4-a_1^4)} ; \qquad 
\nu = \sqrt{ -\frac 1 2 + \lambda^ 2 b^{-2} a_1^4}.
\label{eq:const-large-freq}
\end{equation}
Note that for large $|\lambda| \gg \lambda_M$ these behave as
$$
\rho \sim |\lambda| b^{-1} a_2^2; \qquad 
\mu_\pm  \sim |\lambda| b^{-1} \sqrt{a_2^4-a_1^4}; \qquad 
\nu \sim |\lambda| b^{-1} a_1^2.
$$
Since $a_2 > a_1$ we find that for these large values of $|\lambda|$:
\begin{equation}
\nu < \mu_\pm < \rho ; \qquad \rho-\nu < \mu_\pm.
\label{eq:const-ineq}
\end{equation}

The coefficients $\alpha_\lambda$ and $\beta_\lambda$ are now oscillating, in fact
\begin{align}
  \frac{\beta_\lambda}{\alpha_\lambda} &= \frac
       {B_{+}(\lambda) e^{2 i \nu b T}  + B_{-}(\lambda)  e^{-2 i \nu b T} }{A_{+}(\lambda)  e^{2 i \nu b T} + A_{-}(\lambda) e^{-2 i \nu b T} }
       \label{eq:betaalpha-exact}
\end{align}
where in terms of the variables $\rho,\mu_\pm$ and $\nu$ we have (ignoring the common factor $e^{i \pi (\delta+\gamma)/2}$):
   \begin{align*}
       A_{\sigma}(\lambda) &= \frac{\Gamma(1-2 i \sigma \nu) \Gamma(-2 i \sigma \nu ) \Gamma(1-2i \rho) \Gamma(-2 i \rho) }
{\Gamma( \frac 12 - i \rho  - \mu_-- i \sigma \nu ) \Gamma (\frac 12 -i \rho +\mu_- -i \sigma \nu ) \Gamma( \frac 12 -i \rho- i \mu_+ - i \sigma \nu) \Gamma (\frac 12  -i \rho+i \mu_+ -i \sigma \nu ) },
\intertext{and}
B_\sigma(\lambda) &= \frac{\Gamma(1-2 i \sigma \nu) \Gamma(-2 i \sigma \nu ) \Gamma(1-2i \rho) \Gamma(2i \rho)   }
    {\Gamma( \frac 12 -i \rho-\mu_--i \sigma \nu ) \Gamma (\frac 12 -i \rho+\mu_--i \sigma \nu ) \Gamma( \frac 12+i \rho-i \mu_+- i \sigma \nu )\Gamma (\frac 12 +i \rho  +i \mu_+- i \sigma \nu) }.
\end{align*}

\begin{figure}
  \subfigure[]{\includegraphics[scale=.6]{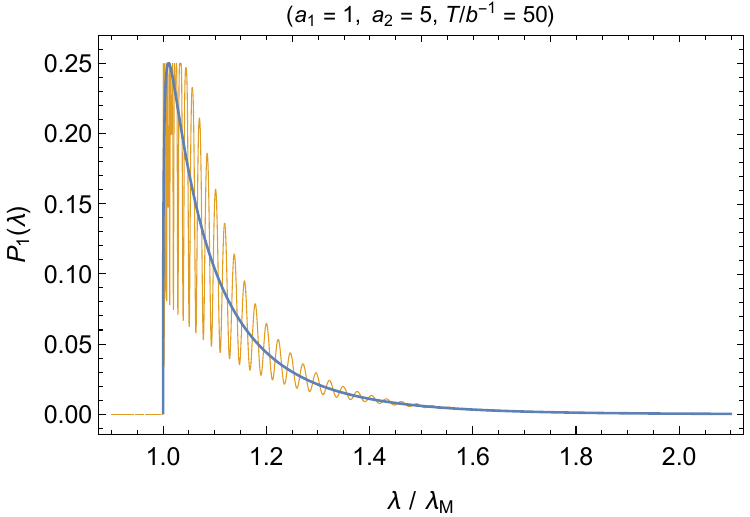}}
  \subfigure[]{\includegraphics[scale=.6]{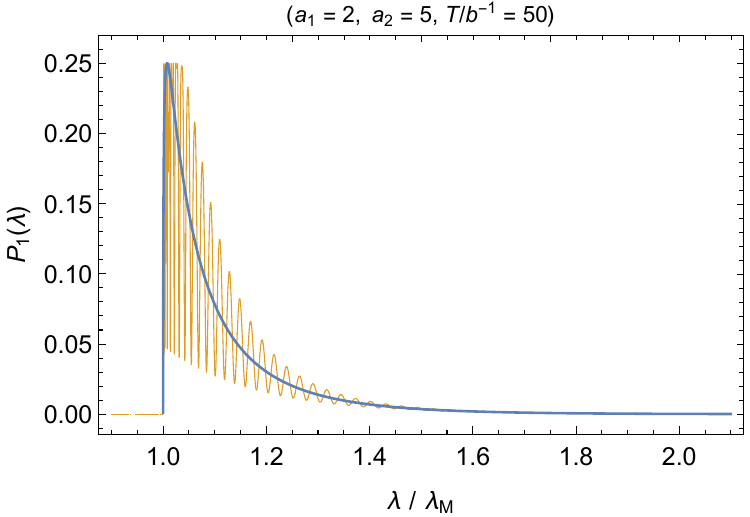}}
  \subfigure[]{\includegraphics[scale=.6]{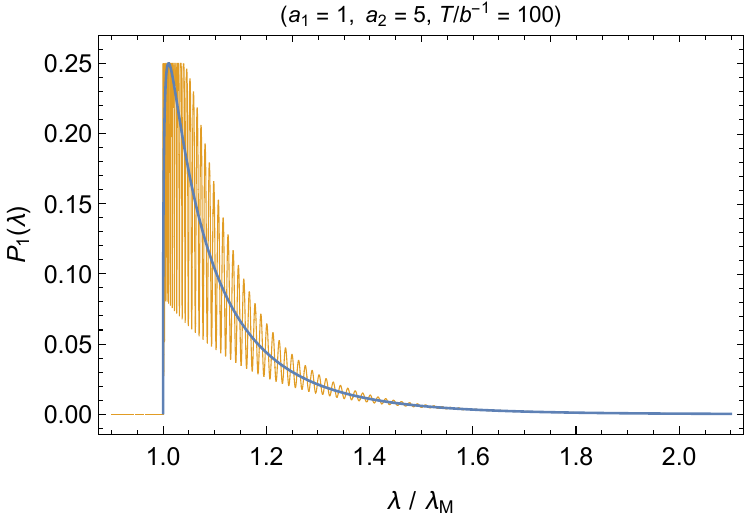}}
  \subfigure[]{\includegraphics[scale=.6]{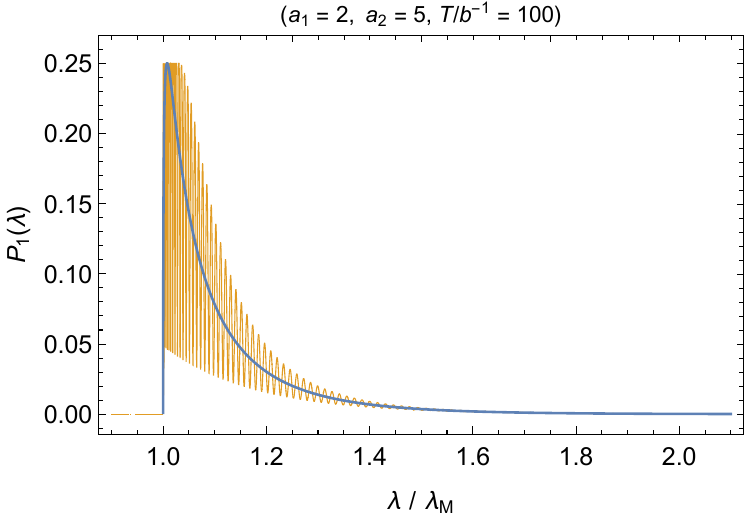}}
  \caption{Probability $P_1(\lambda)$ defined in \eqref{eq:prob-n} in terms of \eqref{eq:betaalpha-exact} for $|\lambda| \geq \lambda_M$ --- for $(a_1=1,a_2=5)$ at different time intervals $T/b^{-1} = 50,100$ in (a), (c) and for $(a_1=2,a_2=5)$ and $T/b^{-1} =50, 100$ in (b), (d). The thick blue line is the approximation of $P_1(\lambda)$ in terms of large $\lambda$-behavior of $|\beta_\lambda|^2/|\alpha_\lambda|^2$ as obtained in \eqref{eq:betaalpha-approx}.}
  \label{fig:prob}
\end{figure}

Let us analyze the leading contributions from the coefficients $A_\pm$ and $B_\pm$ in the limit of large $|\lambda|$, {\em i.e.} for $|\lambda| \gg \lambda_M$. Then we have that the variables $\rho,\mu_+$ and $\nu$ that appear in \eqref{eq:imagin-var} and \eqref{eq:const-large-freq} are large. We will use Stirling's approximation for the gamma-function for large $z$, {\em i.e.}
$$
\Gamma(z) \sim \sqrt{2\pi}  ~ z^{z-1/2} e^{-z} = \sqrt{2\pi}  ~ e^{(z-1/2) \mathrm{Log} ~z - z}; \qquad (z \to \infty : |\arg z| < \pi).
$$
Here, the complex logarithm is given by 
$$
\mathrm{Log}~ z =\log |z| + i \arg z; \qquad (\forall z : |\arg z| < \pi).
$$
Using this, we may analyze the leading behavior for the first two terms in the denominators of $A_\sigma$ and $B_\sigma$. In fact, we have for sufficiently large $a_2 \gg a_1$ that $\rho\sim\mu_-$ which dominates $\nu$, so that $\arg(  \tfrac 12 + \mu_- \pm i (\rho  + \sigma \nu ) \sim \pm \pi /4$ while $\arg(  \tfrac 12 - \mu_- \pm i (\rho  + \sigma \nu ) \sim \pm 3 \pi /4$. As a consequence, we have 
\begin{align*}
  \Gamma( \tfrac 12 + \mu_- + i( \rho  + \sigma \nu )) & \sim \sqrt{2\pi} ~ e^{ \mu_- +i ( \rho + \sigma \nu) \left( \log | \frac 12 + \mu_- + i( \rho  + \sigma \nu) | + i \frac { \pi }{4} \right)  - \left(\frac 12 + \mu_- + i( \rho + \sigma \nu) \right)},\\
  \Gamma( \tfrac 12 + \mu_- - i( \rho  + \sigma \nu )) & \sim  \sqrt{2\pi}~e^{ \mu_- -i ( \rho + \sigma \nu) \left( \log | \frac 12 + \mu_- - i( \rho  + \sigma \nu) | - i \frac { \pi }{4} \right)  - \left(\frac 12 + \mu_- - i( \rho + \sigma \nu) \right)},\\
  \Gamma( \tfrac 12 - \mu_- + i( \rho  + \sigma \nu )) & \sim  \sqrt{2\pi}~e^{- \mu_- +i ( \rho + \sigma \nu) \left( \log | \frac 12 - \mu_- + i( \rho  + \sigma \nu) | + i \frac {3 \pi }{4} \right)  - \left(\frac 12 - \mu_- + i( \rho + \sigma \nu) \right)},\\
    \Gamma( \tfrac 12 - \mu_- - i( \rho  + \sigma \nu )) & \sim  \sqrt{2\pi}~e^{ -\mu_- -i ( \rho + \sigma \nu) \left( \log | \frac 12 - \mu_- -i( \rho  + \sigma \nu) | - i \frac {3 \pi }{4} \right)  - \left(\frac 12 - \mu_- - i( \rho + \sigma \nu) \right)}.\\
  \end{align*}
When combined, this yields for the first two terms in the denominator of $A(\sigma),B(\sigma)$: 
$$
|\Gamma( \tfrac 12 - \mu_- - i \rho  - i \sigma \nu ) \Gamma (\tfrac 12+\mu_- -i \rho  -i \sigma \nu )|^ 2 \sim \left( \frac{2 \pi}{e}\right)^ 2 e^{-2 \pi  (\rho+\sigma \nu)}.
$$
Combining this with the following formulas for the absolute values of the gamma-functions:
\begin{align}
  |\Gamma(i x)|^2&= \frac{\pi}{x \sinh \pi x} \sim \frac{2\pi }{x} ~e^{-\pi x} \qquad (x \to \infty); \nonumber \\
  |\Gamma(\frac 12 + i x)|^2 &= \frac{\pi}{\cosh \pi x} \sim  2\pi  ~e^{-\pi x} \qquad (x \to \infty);
\label{eq:gamma-cosh}
  \\ 
    |\Gamma(1+i x)|^2&= \frac{\pi x}{ \sinh \pi x} \sim 2\pi x ~e^{-\pi x} \qquad (x \to \infty),\nonumber
\end{align}
we find for $A_\sigma$ that
$$
|A_\sigma(\lambda)|^2 \sim e^{2} \frac{e^{-4 \pi\sigma \nu} e^{-4 \pi \rho}}{e^{-2\pi ( \rho +\sigma \nu)} e^{-\pi(\rho+\mu_+ +\sigma\nu)} e^{-\pi \sigma (\rho-\mu_+ + \sigma \nu)}} = e^2 e^{\pi (-1+\sigma) (\rho- \mu_++\nu(4+\sigma))} 
$$
so that
$$
|A_{+}(\lambda)|^2 \sim 1; \qquad |A_{-}(\lambda)|^2 \sim e^{-2\pi(\rho-\mu_+ + 3\nu)}
$$
where we note that $\rho-\mu_+ + 3 \nu > 0$ so that $A_{-}(\lambda)$ is suppressed for large $|\lambda|$ and $A_{+}(\lambda)$ is the dominant term.

Similarly, for $B_\sigma$ we find
$$
|B_\sigma(\lambda)|^2 \sim e^2 e^{\pi(\mu_+ - \rho)(1+\sigma) + \nu (-4+\sigma+\sigma^2)}
$$
so that now
$$
|B_{+}(\lambda)|^2 \sim e^{2 \pi (\mu_+ - \rho-\nu)} ; \qquad |B_{-}(\lambda)|^2 \sim e^{-4 \pi \nu}.
$$
From $\mu_+ - (\rho-\nu) >0$ it follows that $(\mu_+ - \rho-\nu)> -2 \nu$ from which we conclude that the dominant term is $B_{+}(\lambda)$.

Thus, for large $|\lambda|$ the terms $A_{-}(\lambda)$ and $B_{-}(\lambda)$ can be neglected when compared to $A_{+}(\lambda)$ and $B_{+}(\lambda)$. In conclusion, using \eqref{eq:gamma-cosh} we have for large $|\lambda|$:
\begin{align}
\frac{|\beta_\lambda|^2}{|\alpha_\lambda|^2} \sim
\frac{|B_+(\lambda)|^2}{|A_+(\lambda)|^2} &= \frac{|\Gamma( \frac 12+i \rho+i \mu_++ i  \nu )\Gamma (\frac 12 +i \rho  -i \mu_++i  \nu) |^2} {|\Gamma( \frac 12-i \rho+i \mu_++ i  \nu )\Gamma (\frac 12 +i \rho  +i \mu_+- i  \nu) |^2} \nonumber \\
& = \frac{\cosh \pi (-\rho+\mu_++\nu) \cosh \pi(\rho+\mu_+-\nu) }{\cosh \pi (\rho+\mu_++\nu) \cosh \pi (\rho-\mu_++\nu) } 
\label{eq:betaalpha-approx}
\end{align}
In view of Equation \eqref{eq:const-ineq} all four arguments of the $\cosh$-functions are positive, so that the quotient behaves as $e^{2\pi (-\rho+\mu_+-\nu)}$ where $-\rho + \mu_+ - \nu < 0$; this gives an exponential decay for the particle production. Upon writing this in terms of the late time frequency $\omega_\lambda = |\lambda|/a_2$ we find
\begin{equation}
  P_n(\lambda) = e^{-n \beta \omega_\lambda}\left(1-e^{-\beta \omega_\lambda}\right); \qquad \beta =2 \pi \left( a_2^2 + a_1^2 -\sqrt{a_2^4 -a_1^ 4}  \right)a_2 b^{-1} .
\label{eq:Pn}
\end{equation}
Also, the expected number $n(\lambda)$ of particles in mode $\lambda$ ({\em cf.} Eq. \eqref{eq:n-lambda}) is then
$$
n(\lambda)\sim  \frac 1 {e^{\beta \omega_\lambda} - 1}.
$$
We have plotted the probability $P_1(\lambda)$ in Figure \ref{fig:prob}, both in terms of the exact expression \eqref{eq:betaalpha-exact} for $|\beta_\lambda|/|\alpha_\lambda|^2$, as well as for the approximation just obtained in \eqref{eq:betaalpha-approx}.

\subsection{Physical implications}
Let us conclude this section with a few words on the physical implications and possible interpretation of the results just obtained.

We have found in the FLRW-model defined in \eqref{eq:scale-FLRW} that in the limit $T \to \infty$ there is a lower threshold for finite particle production at eigenvalue $\lambda_M = b/ (\sqrt{2} a_1^2)$. In terms of late time frequency $\omega_\lambda = |\lambda|/a_2$ this means that particles are produced with frequency greater than $b/ (\sqrt{2} a_1^2 a_2)$. Interestingly, as we will show next, the reciprocal of this frequency threshold is approximately the duration of the scale change. A physical interpretation of this relation could be that the spatial change is taking place too fast compared to the frequency for an effect to happen.

Let us then derive the claimed relation between the frequency threshold and the duration of the scale change. 
Consider a first approximation of $\a(t)$ given by the linear interpolation $a_{\text{lin}}(t)$ between the scales $a_1$ and $a_2$ for a time interval $[t_0, t_0+\Delta t]$ of length $\Delta t$, {\em i.e.}
\begin{equation}
  a_{\text{lin}}(t) = \frac{a_2-a_1}{\Delta t} (t-t_0)+ a_1.
  \label{eq:a-lin}
\end{equation}
In this case, using Eq. \eqref{eq:tau}
$$
\Delta \tau \sim \int_{t_0}^{t_0+\Delta t} a_{\text{lin}}^{-3}(t) dt 
=\frac{1}  {2(a_2-a_1)}\left( \frac{1}{a_1^2} - \frac{1}{a_2^2}  \right)\Delta t \sim \frac {\Delta t } {2 a_1^2 a_2}
$$
for $a_2 \gg a_1$. Given this linear approximation of $\a(t)$, a time interval $\Delta \tau = 1/b$ thus corresponds to $\Delta t= 2 a_1^2 a_2/b$ as claimed above.

Looking at the profile for particle production \eqref{eq:Pn} we observe that in the spatially flat case this corresponds for large frequencies to a Planckian frequency distribution. Indeed, in the infinite volume limit we may replace the late time frequency $\omega_\lambda$ by a continuous variable $k$, so that the average particle density becomes:
$$
\langle \mathcal N \rangle = (2 \pi^2 a_2^3 )^{-1} \int_0^\infty dk ~ k^2 \left( e^{\beta k }- 1\right)^{-1}.
$$
where the inverse temperature $\beta$ obtained from \eqref{eq:Pn}  is $\beta \sim 2 \pi a_1^2 a_2/b$ for $a_2 \gg a_1$. 

As far as the late time state is concerned, it results from a Bologyubov transformation applied to the vacuum state, and as such it is a coherent superposition of (pairs of) particles, in contrast to blackbody radiation which is actually a highly mixed state. However, as argued at the end of \cite[Section 2.9]{PT09}, local observations would be unable to distinguish between this superposition and the mixed state~\cite{Par76,Par77}.

\section{Outlook}
\label{sect:outlook}
We have considered particle creation both in the case of a long background electric field pulse and a gravitational pulse in an FLRW-spacetime. We have derived probabilities for particle production in terms of the pertinent Bogolyubov coefficients, which were expressed in terms of connecting formulas for the hypergeometric functions and the Heun functions, respectively. In that sense, at the conceptual level the two cases are very similar, however, the resulting outcomes in terms of the probability amplitudes differ. Let us summarize these differences here.

For the electric field pulse, we stress that the coupling to the scalar field is via the electromagnetic vector potential, which has a $\tanh$-profile. Also there is spatial anisotropy, since the electric field is a non-zero constant only in one the three directions.
Concerning the output for the electric field, the resulting particle production can be expressed in terms of the 3-momenta, while the slope parameter $b$ appears in terms of the duration of the interval.

For the gravitational field pulse, the coupling is via the scale function, and is isotropic in the spatial directions. In fact, the whole spatial geometry is scaled uniformly. Nevertheless, from the Heun equation we derive finite particle production to happen with a lower frequency threshold, above which it can be approximated by a Planckian frequency spectrum.  Note that a similar relation between particle production and transition time has been found in the context of de Sitter space, see \cite{AMS18} and the further analysis in \cite{For21}.

In order to further elucidate the role that is played by the time interval of scale change for particle production in the FLRW-model considered in Section \ref{sect:flrw}, we include in Appendix \ref{sect:app-cosh} the analogous derivation for another FLRW-spacetime. The corresponding scale function describes an adiabatic change of scale, and, in fact, the limit of a long gravitational field pulse limit coincides with the adiabatic limit of slow change. We find that in that limit particle production is suppressed. This is further support for the claim that there is a close relation between the particle production and the time interval of scale change. In contrast, Appendix \ref{sect:app-parker-double} provides a model with a short time interval of scale change as compared to the interval of the pulse. This suggests that the exact shape of the pulse does not matter that much, in line with the earlier findings in \cite[Section 2.3-2.4]{PT09}. Nevertheless, it would be interesting to study other types of pulses, and, for instance, the availability of an expansion of the connection coefficients for $a \sim 1$ (instead of the currently expoited $a \sim 0$) would correspond to an extremely short pulse, which is again of physical interest. We leave this for further future research.

Let us compare the results of this paper to our previous works \cite{WSF23,FWS25} where we applied heat kernel methods to derive particle production in a gravitational background. There, the use of Euclidean methods corresponds to the consideration of a thermal equilibrium state, the so-called Hartle--Hawking state. Instead, in the present work we consider the vacuum state in FLRW (at scale $\a(\pm \infty)=a_2$) but by an observer in a spacetime at scale $a_1$. The formalism is different and uses canonical quantization, and Bogolyubov transformations between the early time and late time vector states; this makes it difficult to compare these results directly.  However, a full analysis and physical consideration of these two approaches goes beyond the scope of the present work and we leave it for future study.

One other interesting aspect of our approach with a gravitational pulse width going to infinity is that it resembles a static situation, where, however, the radiation comes from the switch-on and switch-off  of the scale factor change in a finite time interval in the infinite past or future.  In a similar vein, it would be interesting to consider the case of gravitational pair production as in \cite{WSF23,FWS25} for black holes, e.g. Hawking radiation, and potentially other stellar remnants, to see which role the switch-on of the matter configuration and the potential appearance of apparent horizons plays in this case. After all, there is no truly static spacetime in the real universe and every object needs to form first---often on time scales much faster than their lifetimes.

\appendix
\section{Particle creation in long gravitational pulse}
\label{sect:app-cosh}
We may also consider a scale function in an FLRW-spacetime which is similar to the expression considered in the context of the Schwinger effect in \ref{sect:schw}. In contrast to the scale function considered in \ref{sect:heun} there is only one parameter $b$ that controls both the characteristic length of the time interval, {\em i.e.} $[-1/b,1/b]$, and the rate of change (which is proportional to $b$).
Concretely, we may take similar to \eqref{eq:AE}: 
\begin{equation}
  \a(\tau) = \left( a_2^4 - \frac{\left( a_2^4- a_1^4\right)} {\cosh^2 (b \tau/2) }\right)^{1/4},
  \label{eq:scale-FLRW-1}
\end{equation}
which changes scale from $a_2$ at $-\infty$ then exponentially decreases to $a_1$, and then returns to $a_2$ at $+ \infty$. Note that in the limit $b \to 0$ the time interval becomes infinitely long, while the rate of change becomes adiabatic. 

Let us then analyze the differential equation \eqref{eq:eom-tau} for the above $\a(\tau)$. A useful coordinate transformation is $u = e^{b\tau}$ so that it becomes:
\begin{equation}
  \psi_\lambda '' + u^{-1} \psi_\lambda' + \lambda^2 b^{-2} \left( \frac{ a_2^4}{u^ 2} -\frac {4( a_2^ 4-a_1^ 4)} {u (1 + u)^2} \right) \psi_\lambda = 0 ,
  \label{eq:eom-u-1}
\end{equation}
where the prime denotes a derivative with respect to $u$.

Let us analyze the behavior of the differential equation \eqref{eq:eom-u-1} at the singular points $u=0$ and $u=-1$.
At $u =0$ the leading behavior is then
$$
\psi_\lambda'' + u^{-1} \psi_\lambda ' +  \lambda^2 b^{-2}  u^{-2} a_2^4 \psi_\lambda = 0; \qquad ( u \sim 0)
$$
with solutions near $u=0$ given by $\psi_\lambda \sim u^{\pm c}$ where $c =i |\lambda|b^{-1} a_2^ 2$. In order to account for the positive frequency condition at $u = 0$ we take $\psi_\lambda \sim u^{-c}$. 

At $u=-1$ we find leading behavior
$$
\psi_\lambda''  + 4 \lambda^2 b^{-2} (a_2^4 - a_1^4) \frac 1{(1 + u)^2}  \psi_\lambda = 0; \qquad ( u \sim -1)
$$
with solutions given by $\psi_\lambda \sim (1 + u)^{d}$ where
$$
d (d -1 ) =- 4 \lambda^ 2b^{-2} (a_2^4 - a_1^4) .
$$
This suggests the following Ansatz for solutions of \eqref{eq:eom-u-1} of the form
$$
\psi_\lambda (u) = u^{-c} (1+u)^{d} h(u)
$$
where then $h(u)$ is found to satisfy the following differential equation:
\begin{equation}
  h'' (u) +   \left( \frac {1-2c} u + \frac {2d} {1+u} \right) h' (u) + \frac{ (d-2c)d  }{u(1+u) } h(u) = 0.
  \label{eq:ode-h-long_hypergeo}
\end{equation}
Again, this is a hypergeometric equation and a solution of \eqref{eq:eom-u-1} is given by 
$$
\psi_\lambda (u) = u^{-c} (1+u)^{d}  {}_2 F_1 (d, -2c+d,1-2c;-u).
$$
The properties of the hypergeometric function ensure that for this choice, $\psi_\lambda$ behaves as $u^{-c_-} $ at $u\sim 0$, {\em i.e.} the positive frequency solution at $\tau \to -\infty$.

Similar to the derivation of \eqref{eq:deriv-hyperg}, the connecting formulas \eqref{eq:corr-hyperg} for the hypergeometric function can be applied to relate the solution at $u=0$ to the ones at $u=\infty$. In the present case, this yields:
\begin{align*}
  \psi_\lambda \sim \frac{\Gamma(1-2c) \Gamma(-2c) }{\Gamma(-2c+d) \Gamma (1-2c-d) }u^{-c}  + \frac{\Gamma(1-2c) \Gamma(2c) }{\Gamma(d) \Gamma (1-d) }u^{c} =: \alpha_\lambda u^{-c} + \beta_\lambda u^c,
\end{align*}
in terms of the respective positive and negative frequency modes as $\tau \to + \infty$. The ratio of these Bogolyubov coefficients can now be computed and we find for their absolute value: 
$$
{\left|\frac{\beta_\lambda}{\alpha_\lambda}\right|}^2 
= \frac{\sin^2 \pi d}{\sin^ 2 \pi d - \sin ^ 2 2 \pi c },
$$
where we recall $c = i |\lambda| b^{-1} a_2^2$ and $d = \frac 12 + \sqrt {1/4- 4 (a_2^ 4-a_1^ 4) \lambda^2 b^ {-2}}$.
For large $|\lambda|$ we have $d \sim 1/2+ 2i |\lambda| b^{-1} \sqrt{a_2^4-a_1^4}$ so that this behaves as
$$
{\left|\frac{\beta_\lambda}{\alpha_\lambda}\right|}^2  
\sim \frac{\cosh^2 \left(2 \pi |\lambda|  b^{-1} (\sqrt{a_2^4-a_1^4})\right)}{\cosh^2\left(2 \pi |\lambda|  b^{-1} (\sqrt{a_2^4-a_1^4})\right) + \sinh^2 \left(4 \pi |\lambda| b^{-1} a_2^2\right) }
\sim e^{-4 \pi |\lambda|  b^{-1} (a_2^2- \sqrt{a_2^4-a_1^4})} 
$$
If we write this as $\exp(-\beta \omega_\lambda)$ in terms of inverse temperature $\beta$ and the late time frequencies $\omega_\lambda = |\lambda|/a_2$, we conclude that the probabilities $P_n(\lambda)$ in \eqref{eq:prob-n} behave as a Planckian frequency spectrum with $\beta = 4 \pi  a_2(a_2^2- \sqrt{a_2^4-a_1^4})  b^{-1}$. In fact, we also have for large $|\lambda|$ that
$$
n(\lambda) = | \beta_\lambda|^2 \sim e^{-\beta \omega_\lambda}.
$$
In contrast to the Schwinger effect, as well as to the FLRW-model considered in \ref{sect:flrw}, we find that in the limit of a long gravitational pulse given by \eqref{eq:scale-FLRW-1} for $b \to 0$ ---which is also the adiabatic limit--- the particle production is exponentially damped.

\section{Particle creation in a `bouncing' universe}
\label{sect:app-parker-double}
We may also consider a universe where a collapse is followed by an expansion. Recall that a collapse from $a_2$ to $a_1$ is described by the scale function
\begin{equation}
\a(u) = \left(
a_2^4 + u \left [  (a_1^4-a_2^4) (u+1)+\gamma \right ] (u+1)^{-2}
\right) ^{1/4}
\label{eq:scale-PT}
\end{equation}
where as before $u = e^{b \tau}$ and $b,\gamma$ are some parameters. We glue the $\tau$-interval $[-\infty,\infty]$ (labeled by I) to another such interval (labeled by II), again from $[-\infty,\infty]$ but now with the scale function $a(u^{-1})$. In other words, the scale function in the second interval is given by the same formula $a(u)$  but with $a_1$ and $a_2$ interchanged. This describes an expansion from $a_1$ to $a_2$ which is the exact mirror image of the previous collapse. We will exploit this symmetry in deriving and then composing the Bogolyubov transformations. 

The scale function in \eqref{eq:scale-PT} for interval I has been considered in \cite[Section 2.8]{PT09}, where it was shown that initial positive frequency solutions $\psi_\lambda^+ \sim u^{-c_2}$ (at $u \sim 0$) with $c_2 = i |\lambda| b^{-1} a_2^2$ evolve to the solution  
$$
\psi_\lambda^+ \sim \alpha^I_\lambda u^{-c_1}  + \beta^I_\lambda u^{c_1}, \qquad (u \sim \infty),
$$
where now $c_1 = i |\lambda| b^{-1} a_1^2$. 
The Bogolyubov coefficients are given by Eq. (2.114) in {\em loc. cit.}:
\begin{align*}
  \alpha^I_\lambda &= \left( \frac {a_1}{a_2} \right) \frac{ \Gamma(-2c_1) \Gamma(1-2c_2)}{\Gamma(d-c_1-c_2)\Gamma(1-c_1-c_2-d)}\\
  \beta^I_\lambda &= \left( \frac {a_1}{a_2} \right) \frac{ \Gamma(2c_1) \Gamma(1-2c_2)}{\Gamma(d+c_1-c_2)\Gamma(1+c_1-c_2-d)}.
  \end{align*}
where $d$ solves $d (d-1) =\gamma \lambda^2/b^2$. 

Interestingly, if we would have started with an initial {\em negative} frequency solution $\psi_\lambda^- \sim u^{c_2}$ we can follow the same computation with $c_2$ replaced by $-c_2$, so that 
$$
\psi_\lambda^- \sim \gamma^I_\lambda u^{-c_1}  + \delta^I_\lambda u^{c_1}, \qquad (u \sim \infty),
$$
where now
\begin{align*}
  \gamma^I_\lambda &= \left( \frac {a_1}{a_2} \right) \frac{ \Gamma(-2c_1) \Gamma(1+2c_2)}{\Gamma(d-c_1+c_2)\Gamma(1-c_1+c_2-d)}\\
  \delta^I_\lambda &= \left( \frac {a_1}{a_2} \right) \frac{ \Gamma(2c_1) \Gamma(1+2c_2)}{\Gamma(d+c_1+c_2)\Gamma(1+c_1+c_2-d)}.
  \end{align*}

For the interval II the scale function is $\a(u^{-1})$, or, equivalently, it is \eqref{eq:scale-PT} with $a_1$ and $a_2$ interchanged. But then the same applies to the Bogolyubov transformations, to yield that positive/negative frequency solutions $\psi_\lambda^\pm \sim u^{\mp c_1}$ (at $u \sim 0$) evolve to a combination:
\begin{align*}
  \psi^+ \sim  \alpha^{II}_\lambda u^{-c_2}  + \beta^{II}_\lambda u^{c_2}, \qquad (u \sim \infty),\\
  \psi^-\sim \gamma^{II}_\lambda u^{-c_2}  + \delta^{II}_\lambda u^{c_2}, \qquad (u \sim \infty).
\end{align*}
with 
\begin{align*}
  \alpha^{II}_\lambda &= \left( \frac {a_2}{a_1} \right) \frac{  \Gamma(1-2c_1)\Gamma(-2c_2)}{\Gamma(d-c_1-c_2)\Gamma(1-c_1-c_2-d)},\\
  \beta^{II}_\lambda &= \left( \frac {a_2}{a_1} \right) \frac{  \Gamma(1-2c_1)\Gamma(2c_2)}{\Gamma(d-c_1+c_2)\Gamma(1-c_1+c_2-d)}.
  \intertext{and}
  \gamma^{II}_\lambda &= \left( \frac {a_2}{a_1} \right) \frac{ \Gamma(1+2c_1)\Gamma(-2c_2) }{\Gamma(d+c_1-c_2)\Gamma(1+c_1-c_2-d)}\\
  \delta^{II}_\lambda &= \left( \frac {a_2}{a_1} \right) \frac{ \Gamma(1+2c_1)\Gamma(2c_2) }{\Gamma(d+c_1+c_2)\Gamma(1+c_1+c_2-d)}.
  \end{align*} 

\begin{figure}
  \subfigure[]{\includegraphics[scale=.6]{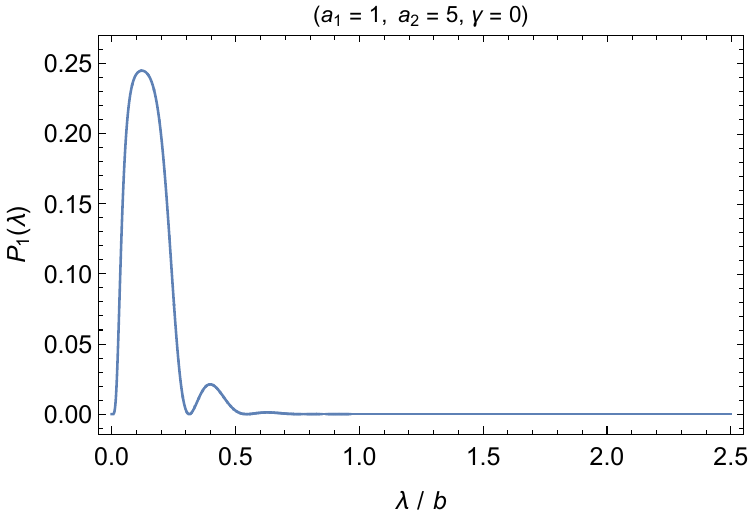}}
  \hspace{5mm}
  \subfigure[]{\includegraphics[scale=.6]{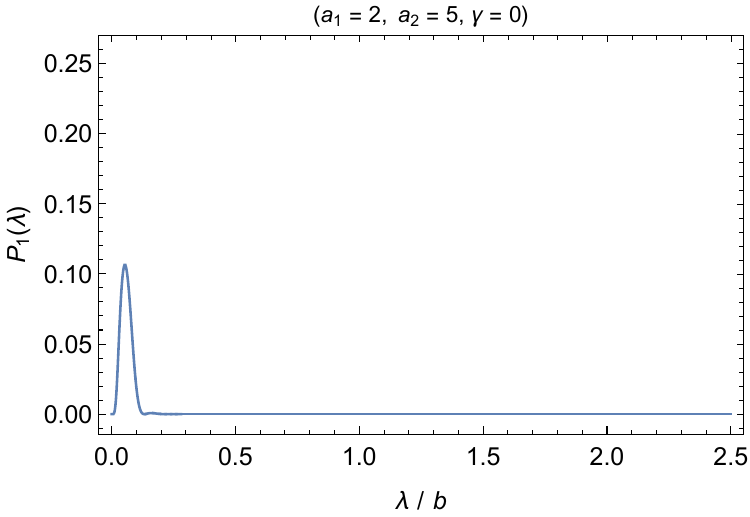}}
    \caption{Probability $P_1(\lambda)$ for $(a_1=1,a_2=5,\gamma=0)$  and $(a_1=2,a_2=5,\gamma=0)$.}
\label{fig:part-prod-bounce}
\end{figure}

The particle creation obtained from this bounce by gluing the intervals I (collapse) and II (expansion) corresponds to the composition of these Bogolyubov coefficients. Indeed, if we start with an initial positive frequency $f_\lambda^+$ at $\tau \to - \infty$ in interval I, this mode becomes a linear combination of positive and negative frequency modes at $\tau \to \infty$ in interval II via:
$$
f_\lambda^+ =\left(  \alpha_\lambda^I  \alpha_\lambda^{II} + \beta_\lambda^I  \gamma_\lambda^{II} \right)  u^{-c_2} + \left( \alpha_\lambda^I \beta_\lambda^{II} + \beta_\lambda^I \delta_\lambda^{II}\right) u^{c_2}  =: \alpha_\lambda u^{-c_2} + \beta_\lambda u^{c_2}.
$$
Note the similarity with the connecting formulas for the Heun equation, which were also sums of two ratios of gamma-functions. The resulting particle production probabilities are depicted in Figure \ref{fig:part-prod-bounce}.

\newcommand{\noopsort}[1]{}

\end{document}